\documentclass[12pt]{article}
\usepackage{graphicx, amsmath,float}
\usepackage{multirow}
\usepackage{booktabs}
\usepackage{array}
\usepackage{lineno}
\usepackage{units}
\usepackage{color}

\newlength{\dinwidth}
\newlength{\dinmargin}
\def\lapproxeq{\lower .7ex\hbox{$\;\stackrel{\textstyle                                                    
<}{\sim}\;$}}                                                    
\def\gapproxeq{\lower .7ex\hbox{$\;\stackrel{\textstyle                                                    
>}{\sim}\;$}}                                                                                                  
\def\bea{\begin{eqnarray}}                                                    
\def\eea{\end{eqnarray}}

\def\sh{\hat s}
\def\sh2{{\hat s}^2}

\newcommand{\be}{\begin{equation}}
\newcommand{\ee}{\end{equation}}

\begin{document}
                                                    
\titlepage                                                    
\begin{flushright}                                            
IPPP/24/41  \\  
LTH 1381 \\                             
\vspace{0.3cm} 

\today \\                                                    
\end{flushright} 
\vspace*{0.5cm}
\begin{center} 

{\Large \bf A method to include exclusive heavy vector-meson \\ production data at small $x$ in global parton analyses}\\
\vspace*{1cm}
C.A. Flett$^{a}$, A.D. Martin$^b$, M.G. Ryskin$^{c}$ and T. Teubner$^d$\\ 

\vspace*{0.5cm}  
\fontsize{10.47}{1}
$^a${\it Université Paris-Saclay,
CNRS, IJCLab, 91405 Orsay, France}\\
$^b${\it Institute for Particle Physics Phenomenology, Durham University, Durham, DH1 3LE, U.K.} \\                         $^c${\it Petersburg Nuclear Physics Institute, NRC Kurchatov Institute, Gatchina, St.~Petersburg, 188300, Russia}  \\
$^d${\it Department of Mathematical Sciences, University of Liverpool, Liverpool, L69 3BX, U.K.}\\

\vspace*{1cm}                                                    

\begin{abstract} 
\vspace*{0.2cm} 
We propose a method which allows the inclusion of exclusive heavy vector-meson production data at low $x$ in future global parton analyses. As an example we perform a study within \texttt{xFitter} to determine the gluon parton distribution function (PDF) at next-to-leading order~(NLO) at moderate-to-low $x$ using the measurements of exclusive $J/\psi$ production in $ep$ and $pp$ collisions from HERA and LHC. 
We further study the constraints from the corresponding $\Upsilon$ production process. We finish by discussing the possible effects at next-to-next-to-leading order~(NNLO) through incorporation of a $K$ factor for the exclusive heavy vector-meson coefficient function at NLO.
\end{abstract}     
\vspace*{0.5cm}                                                                                                
\end{center} 

\section{Introduction}

The precision of the parton distributions of the proton is well established through global analyses, provided the resolution scale $Q^2$ is not too low and the momentum fraction $x$ remains within a moderate range, neither too small nor too large. 
For $x \gapproxeq 10^{-3}$, there is a general agreement among the results from various global fit analyses~\cite{1,2,3}. However, as $x$ decreases, particularly at lower scales, the uncertainty in these distributions increases significantly. This increase in uncertainty is due to the lack of experimental data directly probing this region. 

The experiments at the Large Hadron Collider~(LHC) are capable of particle detection and reconstruction over a wide rapidity range. 
Notably, various measurements of the differential cross sections for exclusive heavy vector mesons such as $J/\psi$ and $\Upsilon$~\cite{LHCb7, LHCb13, LHCb:2015wlx} 
have allowed for the determination of the gluon Parton Distribution Function~(PDF) down to $x \sim 3\times 10^{-6}$ at factorization scales $\mu_F = m_q$, where $q = c,b$.

Unfortunately it is not easy to include these data in global parton analyses. The cross section for exclusive meson production is driven not by the usual (diagonal) PDFs but by the 
more complicated (skewed) generalised parton distributions (GPDs),  and is proportional to the skewed gluon-density squared, as indicated in Fig. 1; see~\cite{Diehl} for a review. Note from the caption of the figure that $x \approx 2\xi$ at very small $x$, where $2\xi$ is the proton momentum fraction transferred through the GPD to the vector meson.

\begin{figure} [t]
\begin{center}
\includegraphics[width=0.4\textwidth]{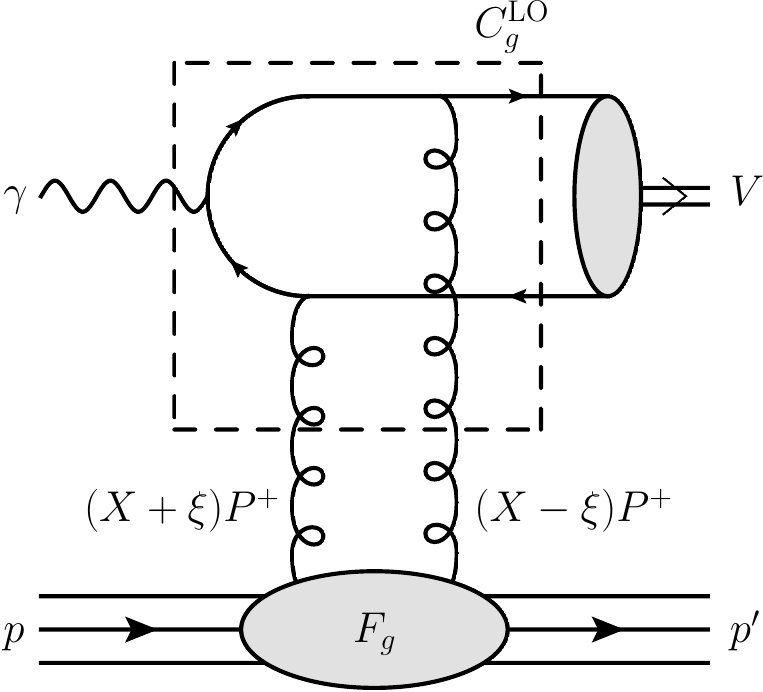}
\qquad
\includegraphics[width=0.4\textwidth]{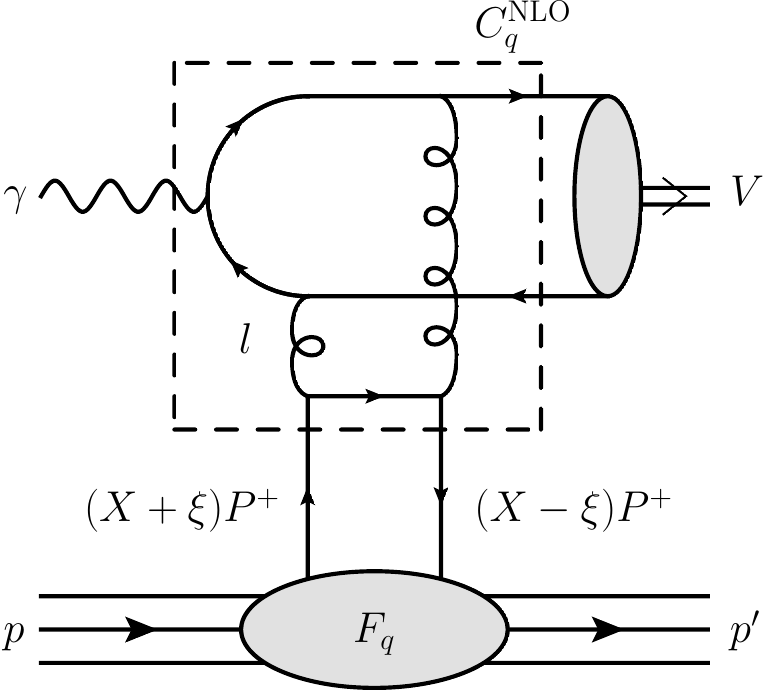}
\caption{\sf{(Left) LO contribution to  $\gamma p \to V +p$, where the vector meson $V = J/\psi, \Upsilon$. (Right)~NLO quark 
contribution. For  these graphs all permutations of the parton lines and couplings of
  the gluon lines to the heavy-quark pair are to be understood. The NLO gluon contribution, with coefficient function $C_g^{{\rm NLO}}$ and GPD $F_g$, dominates the NLO quark contribution. In these diagrams, the momentum  
  $P\equiv (p+p^\prime)/2$ and $l$ is the loop momentum. Note that the momentum fractions of the left and right partons are $x=X+\xi$ and $x'=X-\xi$ respectively; for the gluons connected to the heavy quark-antiquark pair, we have $x' \ll x$ and so $x\simeq 2\xi$.
  }}
\label{fig:f2}
\end{center}
\end{figure}

However, in the small $x$ region of our interest, the value of the GPD can be calculated from the conventional PDF with good ($\sim O(x)$) accuracy using the Shuvaev transform~\cite{Shuv}. The Shuvaev transform makes use of the fact that as $\xi\to 0$ (and transverse-momentum transfer $p_T = 0$) the Gegenbauer moments\footnote{ Gegenbauer moments are the analogue of Mellin moments which diagonalize the $Q^2$ evolution of PDFs. The
corresponding operator diagonalizes the $Q^2$ evolution of the GPDs \cite{Ohr}. As $\xi\to 0$ the Gegenbauer moments
become equal to the Mellin moments.}  of the GPD become equal to the
known Mellin moments of the PDF. Due to the polynomial condition (see e.g. \cite{Ji}) even for $\xi\neq 0$ the Gegenbauer moments can be obtained from the Mellin moments to $O(\xi)$ accuracy.
Thus it is possible to obtain the full GPD function at small $\xi$ from its known moments. 

In principle, this allows us to include the low-$x$ exclusive $J/\psi$ data in the global PDF analysis. In practice the problem is that the Shuvaev transform amounts to a slowly convergent double integral and the corresponding computation is time consuming. This is troublesome for the global fit as after every iteration, a new three-dimensional grid defined over the variables $X, \xi$ and $Q^2$ has to be computed to obtain the updated theory prediction.

In the present paper we describe a method which helps to overcome this problem by replacing the exclusive vector-meson data at small $x$ by `effective gluon points’. To demonstrate the efficiency of the proposed method and to show how the inclusion of the LHCb $J/\psi$ and $\Upsilon$ exclusive cross sections affect the result we compare the standard \texttt{xFitter}~\cite{Alekhin:2014irh} NLO parton analysis of DIS data with that supplemented by the vector-meson data.
Note that our goal is not to present a new set of parton distributions but to emphasise the possible role of the vector-meson data in global parton analyses. The new `effective gluon points’ presented in Tables~\ref{tab:1} and \ref{tab:2} can be used in future global analyses.

\section{Effective low-$x$ gluon PDF data}
To overcome this difficulty, we propose the following procedure.
We translate the experimental $J/\psi$ cross section data into a set of ``effective" values for the gluon PDF.
 
As seen from Fig. 2, the cross section for the exclusive process 
\be 
p+p\to p+ J/\psi+p
\ee
contains two components,
\begin{equation}
    \frac{d\sigma^{(pp \rightarrow p+J/\psi+p)}}{dY}~ = ~S^2(W_+)\left(k_+\frac{dn}{dk_+}\right)\sigma_+(\gamma p \rightarrow J/\psi p)~+~W_-~{\rm term},
    \label{diff1}
\end{equation}
corresponding to energies $W_{\pm}$, where $W_+$ denotes the energy when the vector meson travels in the same direction as the photon, and $W_-$ when it travels in the opposite direction.
We see that the values of the subprocess cross sections $\sigma(\gamma p \to J/\psi p)$, for each rapidity $Y$ of the $J/\psi$, are weighted by the corresponding gap survival factors $S^2(W_\pm)$~\cite{Khoze:2013jsa} and photon fluxes $dn/dk_\pm$~\cite{Budnev:1975poe}. The cross section  with the lower $\gamma p$-energy ($W_-$) was measured at HERA~\cite{ZEUS:2002wfj, ZEUS:2004yeh, H1:2005dtp, H1:2013okq} and can be calculated with sufficient accuracy since it corresponds to relatively large $x$ where the uncertainties in the parton distributions are small. 
Hence the exclusive $J/\psi$ data provide reliable values for $\sigma_+(\gamma+ p\to J/\psi +p)$, corresponding to the component with the larger $\gamma p$-energy, $W_+$, see Fig.~3.\footnote{Actually, the subtraction of the $W_-$ contribution was done already in the LHCb analyses where the results are presented as the $\sigma_+(\gamma + p \rightarrow V+p)$ cross sections. } 

\begin{figure}
\begin{center}
\includegraphics[scale=0.55]{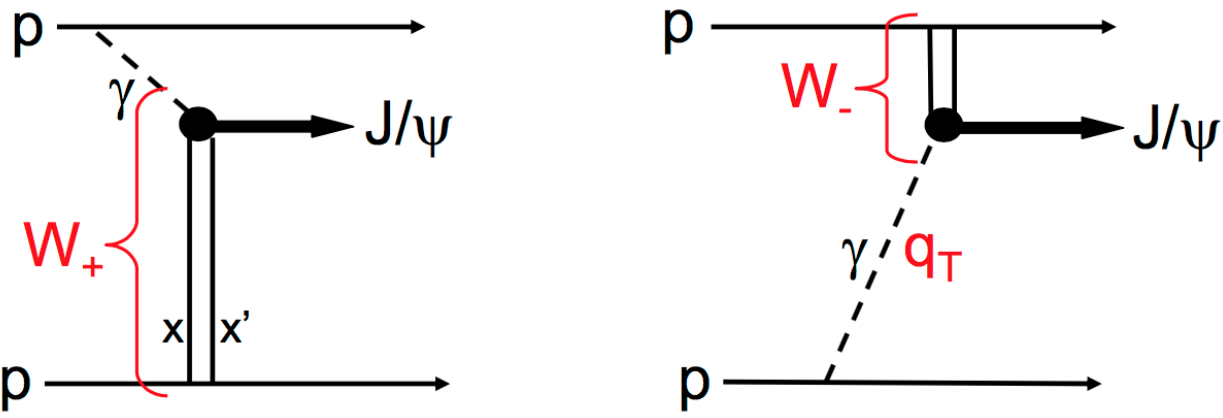}
\caption{\sf{The two diagrams describing exclusive $J/\psi$ production at the
  LHC. The vertical lines represent two-gluon exchange. The left diagram,
  the $W_+$ component, is the major contribution to the $pp \to p+J/\psi
  +p$ cross section for a $J/\psi$ produced at large rapidity $Y$. Thus
  such data allow a probe of very low $x$ values, $x\sim M_{J/\psi}\,{\rm
    exp}(-Y)/\sqrt{s}\,$, where $\sqrt{s}$ is the centre-of-mass energy of the $pp$ system; recall that for two-gluon exchange we have
  $x\gg x'$.   The $q_T$ of the photon is very small and so the photon can be considered as a real on-mass-shell particle.}}    
\label{LHCb}
\end{center}
\end{figure}

 These $\sigma_+$ data were well described in~\cite{prd102} by a power-like low-$x$ gluon
\be
 xg(x)=A\cdot x^{-\lambda} ~~~~~  {\rm with} ~~~~~ \lambda=0.135\pm 0.006,
 \label{gg}
\ee
also shown in Fig. 3.

\begin{figure} [t]
\begin{center}
\includegraphics[scale=0.7]{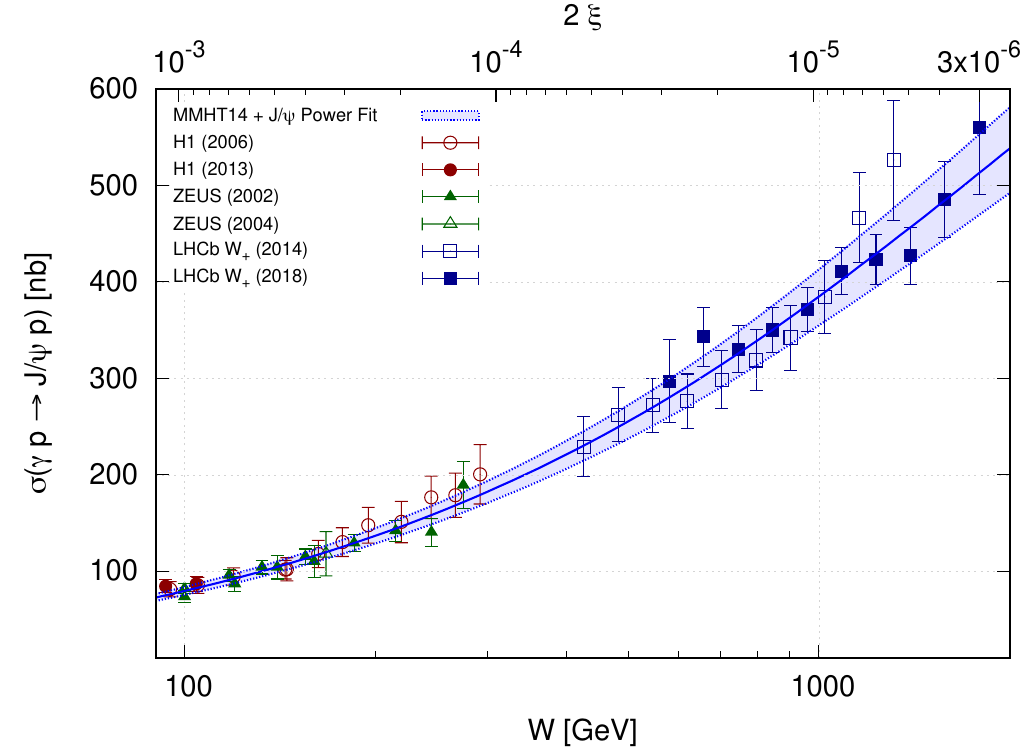}
\caption{\sf{The description of the $J/\psi$ photoproduction HERA~\cite{ZEUS:2002wfj, ZEUS:2004yeh, H1:2005dtp, H1:2013okq} and LHCb~\cite{LHCb7, LHCb13} data based on using the central value of the global gluon PDF from the MMHT14 global analysis for $x>10^{-3}$ and a fitted power-like gluon PDF, see eqn.~\eqref{gg}, for $x<10^{-3}$. 
The blue solid line and shaded region show, respectively, the central prediction and propagation of the $\pm\,1 \sigma$ errors on the fitted parameters to the cross-section level. 
}}
\label{fig:3}
\end{center}
\end{figure}

Moreover it was shown that after the optimal factorisation scale\footnote{At this optimal scale the large double-logarithmic contributions of the form $c_n(\alpha_s\ln(1/\xi)\ln \mu_F^2)^n$ are resummed into the incoming PDF/GPD~\cite{opt}.},
$\mu_F = M_{J/\psi}/2$, was chosen, where $M_{J/\psi}$ is the mass of the $J/\psi$, and the so-called $Q_0$ subtraction~\cite{Q0}\footnote{The $Q_0$ subtraction is necessary to exclude the double counting between the low $l_T$ part of the coefficient function and the PDF-{\em input}. This, together with the resummation of the aforementioned double-logarithmic terms~\cite{opt}, helps mitigate the problem of the strong factorisation scale dependence~\cite{nlo} of the NLO amplitude.} was performed, the quark contribution to $\sigma_+$ becomes negligible~\cite{Flett:2019pux}. That is the value of 
\be
 \sigma_+ \,\propto \,(\,2\xi\, g(2\xi)\,)^2
\ee
is just proportional to the gluon density squared.\footnote{Strictly speaking, the value of $\sigma_+$ is proportional to the square of the corresponding GPD.}
 
 Now for each data point, $i$,
 we calculate the predicted $J/\psi$ cross section $\sigma_+({\rm fit})_i$ based on the gluons from our fit and compare it with the experimental value $\sigma_+({\rm data})_i$. The ratio $\sigma_+({\rm data})_i/\sigma_+({\rm fit})_i$ is equal to the square  of the ratio of the gluon densities
 \be 
 \frac{\sigma_+({\rm data})_i}{\sigma_+({\rm fit})_i}~~=~~\left(\frac{g_{{\rm eff}}(2\xi_i)}{g_{\rm fit}(2\xi_i)}\right)^2.
 \ee 
In other words, in this way we can calculate the effective gluon density 
(PDF) corresponding to $x_i=2\xi_i$ as
\be
g_{{\rm eff}}(x_i,\mu_{{\rm opt}})~~=~~g_{{\rm fit}}(x_i,\mu_{{\rm opt}})~~\sqrt{\frac{\sigma_+{({\rm data})_i}}{\sigma_+({\rm fit})_i}} .
\label{glu}
\ee
Correspondingly the error
\be
\delta g_{\rm_{{\rm eff}}}(x_i)=\frac 12 ~~g_{{\rm eff}}(x_i)~~\frac{\delta\sigma_+({\rm data})_i}{\sigma_+({\rm data})_i} .
\label{err}
\ee

The error of the lower energy contribution, $\sigma_-(x_i)$, is accounted for in
the experimental value of $\sigma_+({\rm data})_i$. Recall that this lower energy contribution is small and
its errors are negligible.

\begin{table}[ht!]
\centering
\begin{tabular}[t]{lcccc}
\toprule
&13 TeV&$x$&$xg_{\rm eff}$&$\delta xg_{\rm eff}$\\
\midrule
&&$2.84 \times 10^{-5}$&3.83&0.26 \\
&&$2.21 \times 10^{-5}$&4.08&0.17\\
&&$1.72 \times 10^{-5}$&3.96&0.14\\
&&$1.34 \times 10^{-5}$&4.05&0.13\\
&&$1.04 \times 10^{-5}$&4.14&0.12\\
&&$8.14 \times 10^{-6}$&4.32&0.13\\
&&$6.34 \times 10^{-6}$&4.35&0.13\\
&&$4.93 \times 10^{-6}$&4.34&0.14\\
&&$3.84 \times 10^{-6}$&4.60&0.17\\
&&$2.99 \times 10^{-6}$&4.91&0.27\\
\midrule
&7 TeV&&&\\
&&$5.28 \times 10^{-5}$&3.45&0.24 \\
&&$4.11 \times 10^{-5}$&3.65&0.20 \\
&&$3.20 \times 10^{-5}$&3.68&0.19 \\
&&$2.49 \times 10^{-5}$&3.67&0.19 \\
&&$1.94 \times 10^{-5}$&3.79&0.19 \\
&&$1.51 \times 10^{-5}$&3.88&0.19 \\
&&$1.18 \times 10^{-5}$&3.99&0.20 \\
&&$9.18 \times 10^{-6}$&4.19&0.20 \\
&&$7.15 \times 10^{-6}$&4.59&0.23 \\
&&$5.57 \times 10^{-6}$&4.84&0.28 \\
\bottomrule
\end{tabular}
\caption{\sf{The values of $xg_{\rm eff}$ and $\delta xg_{\rm eff}$ obtained from eqns.~\eqref{glu} and~\eqref{err} at $Q^2 = M_{J/\psi}^2/4 = 2.4\,{\rm GeV}^2$ using the exclusive $J/\psi$ production data at 7 TeV~\cite{LHCb7} and 13 TeV~\cite{LHCb13} obtained by the LHCb collaboration. 
}}
\label{tab:1}
\end{table}

To summarize:  we propose to include these ``effective" gluon PDF data points calculated according to eqns.~\eqref{glu},~\eqref{err} in our parton analysis in order to reduce the present uncertainties in the behaviour of the low-$x$ partons.

Since the final parton distribution may differ from eqn.~\eqref{gg} obtained from \cite{prd102}, to achieve greater accuracy several iterations may be needed such that the resulting output parameters in iteration $i$ do not differ from those in iteration $i-1$ by $\pm 1 \sigma$. The final determined {\em iterated} data points are given in Table~\ref{tab:1}.\footnote{Note that we do not fit to the individual $\sigma(\gamma+p\to J/\psi+p)$ data points in our approach. Instead, we use effective $xg_{{\rm eff}}(x)$ points. Recall the values of $xg_{\rm fit}(x)$ were obtained from a previous fit to the $J/\psi$ cross section only, and thus may not be sufficiently accurate for a global analysis. Therefore we do not use the values of $xg_{\rm fit}(x)$ from this former $J/\psi$ fit, but instead fit to the corrected $g_{{\rm eff}}(x)=g_{{\rm fit}}(x)\sqrt{\sigma({\rm data})(x)/\sigma({\rm{fit})}(x)}$ data points. These effective values will provide a better overall description of the data and a better stability, so that the true values will not be changed too much in the fit iteration.}

\begin{table}[ht!]
\centering
\begin{tabular}[t]{lcccc}
\toprule
&7, 8 TeV&$x$&$xg_{\rm eff}$&$\delta xg_{\rm eff}$\\
\midrule
&&$1.01 \times 10^{-4}$&19.9&3.1 \\
&&$4.77 \times 10^{-5}$&23.9&3.3\\
&&$2.26 \times 10^{-5}$&31.1&4.1\\
\bottomrule
\end{tabular}
\caption{\sf{The values of $xg_{\rm eff}$ and $\delta xg_{\rm eff}$ obtained from eqns.~\eqref{glu} and~\eqref{err} at $Q^2 = M_{\Upsilon}^2/4 = 22.4\,{\rm GeV}^2$ using the exclusive $\Upsilon$ production data at 7, 8 TeV~\cite{LHCb:2015wlx} obtained by the LHCb collaboration.}} 
\label{tab:2}
\end{table}%

In Table~\ref{tab:2}, we show the final effective iterated gluon points for exclusive $\Upsilon$ production~\cite{LHCb:2015wlx} at the corresponding `optimal' scale 22.4 GeV$^2$.  
The accuracy of these points is worse than that obtained from $J/\psi$ production, but they allow for a constraint on the $Q^2$-evolution in the low-$x$ region.\footnote{So far the exclusive $\Upsilon$ data for the $\gamma p \to \Upsilon p$ cross section from the LHCb collaboration have been obtained  using gap survival factors $S^2(W_{\pm})$ and photon fluxes $dn/dk_{\pm}$ that need updating or correcting, see~\cite{Flett:2021fvo}. Nevertheless, the resulting $\Upsilon$ data shown in Table 2 already include the corresponding corrections,
which however never exceed the experimental error bars.
Note that the resulting $\Upsilon$ points already show evidence of a rising gluon PDF power behaviour in the region $10^{-5} < x < 10^{-4}$ and $Q^2 = m_b^2 \sim 22~{\rm GeV}^2$ (see Fig.~2 of~\cite{Flett:2021fvo}). Recall that as was shown in~\cite{Flett:2021fvo}, at the border of the LHCb acceptance, the choice of survival factor and photon flux only amounted to 5\% for $J/\psi$ production, while for $\Upsilon$ production it is of the order of 25\%.}

For the results presented in Tables~\ref{tab:1} \& \ref{tab:2}, we actually use the Shuvaev transform for $x \leq 10^{-4}$ ($x \leq 10^{-3}$ for the data presented in Fig.~3).

\section{The impact of the low-$x$ gluon effective data points on a global analysis}
\begin{figure} [t]
\begin{center}
\includegraphics[scale=0.38]{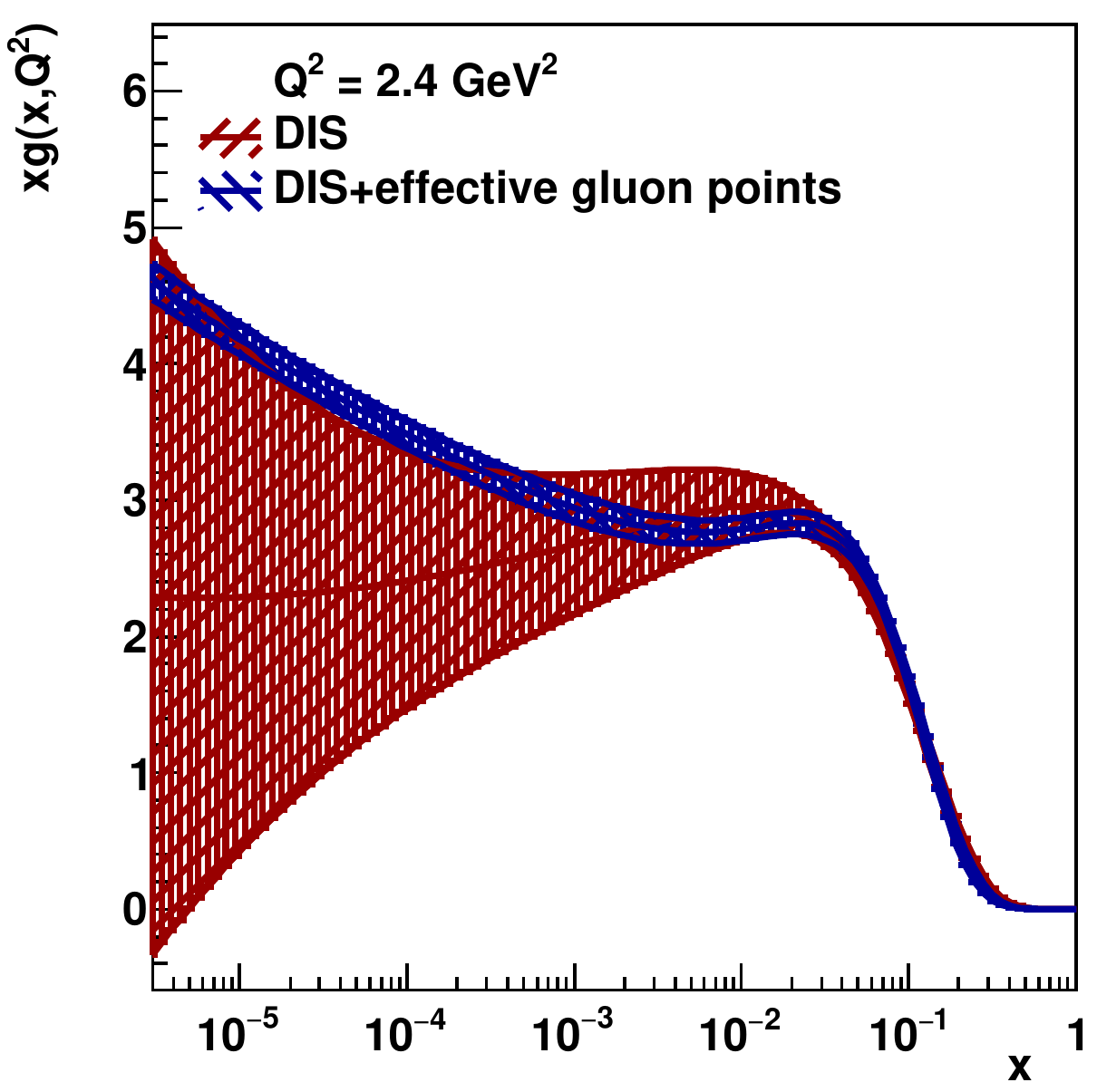}
\includegraphics[scale=0.38]{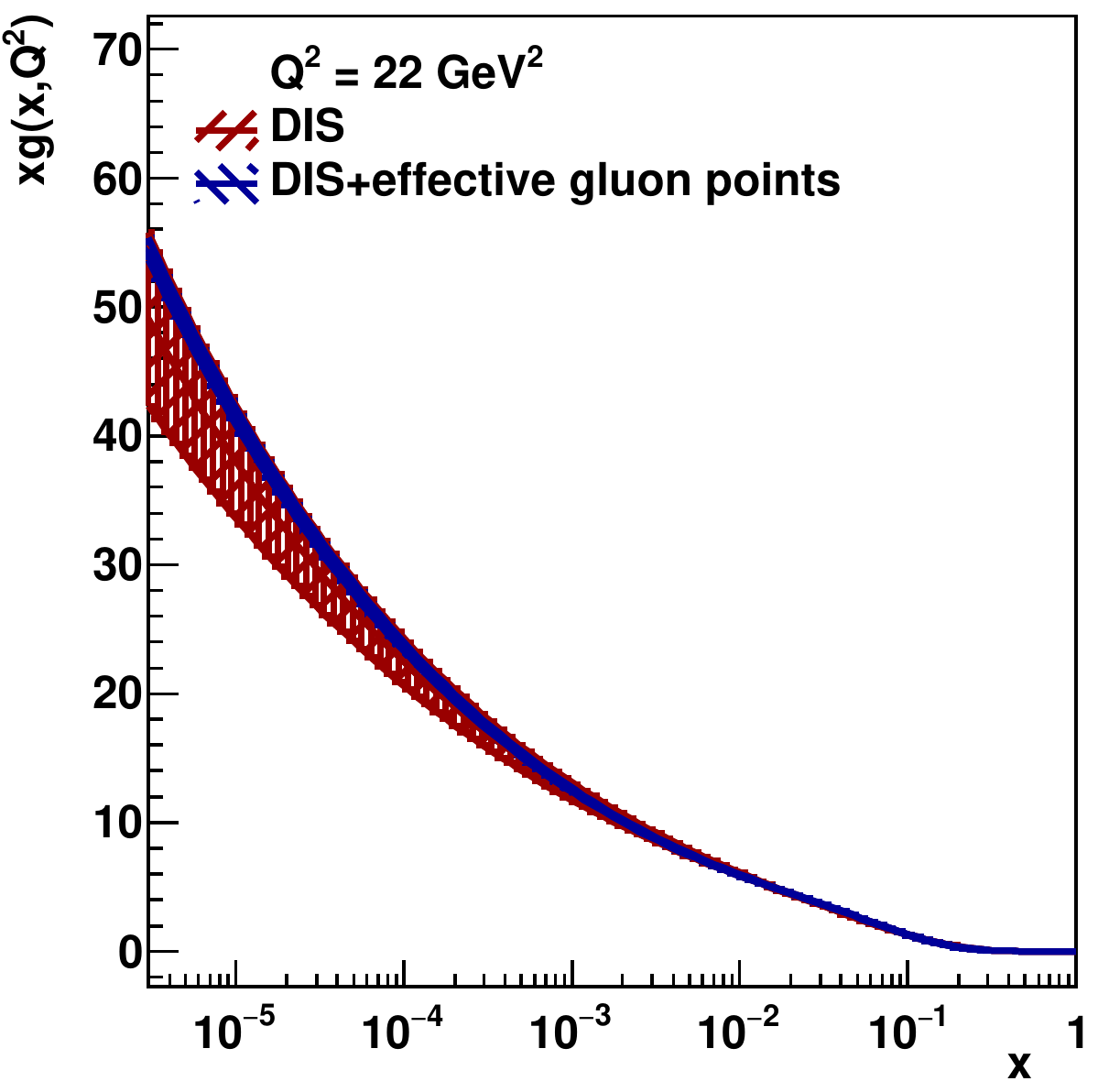}
\caption{\sf{The gluon distributions $xg$ at scales $\mu_F^2 = Q^2 =$ 2.4 GeV$^2$ (left) and 22 GeV$^2$ (right) obtained in \texttt{xFitter} fitting DIS (red) and DIS+effective gluon data points (blue).}}
\label{fig:4}
\end{center}
\end{figure}

The gluon parton distribution resulting from a global \texttt{xFitter}~\cite{Alekhin:2014irh}
NLO analysis of pure Deep-Inelastic-Scattering~(DIS) data (red) and DIS data supplemented with the effective gluon points extracted from the exclusive $J/\psi$ and $\Upsilon$ production at the LHC (blue) are shown in Fig.~\ref{fig:4}. Here we have used the standard \texttt{xFitter} ensemble of DIS data and the standard
\texttt{xFitter} ansatz for the input parton parametrizations with the kinematic cut $Q^2 > 2.4$ GeV$^2$. NLO coefficient functions and NLO PDFs were used to obtain the $g_{\text{fit}}$ values. The datasets used and the partial and total minimum $\chi^2$ figure of merit per degree of freedom, $\chi_{\text{min}}^2/\text{d.o.f}$, for both fits are shown in Table~\ref{tab:3}.

\begin{figure} [t]
\begin{center}
\includegraphics[scale=0.38]{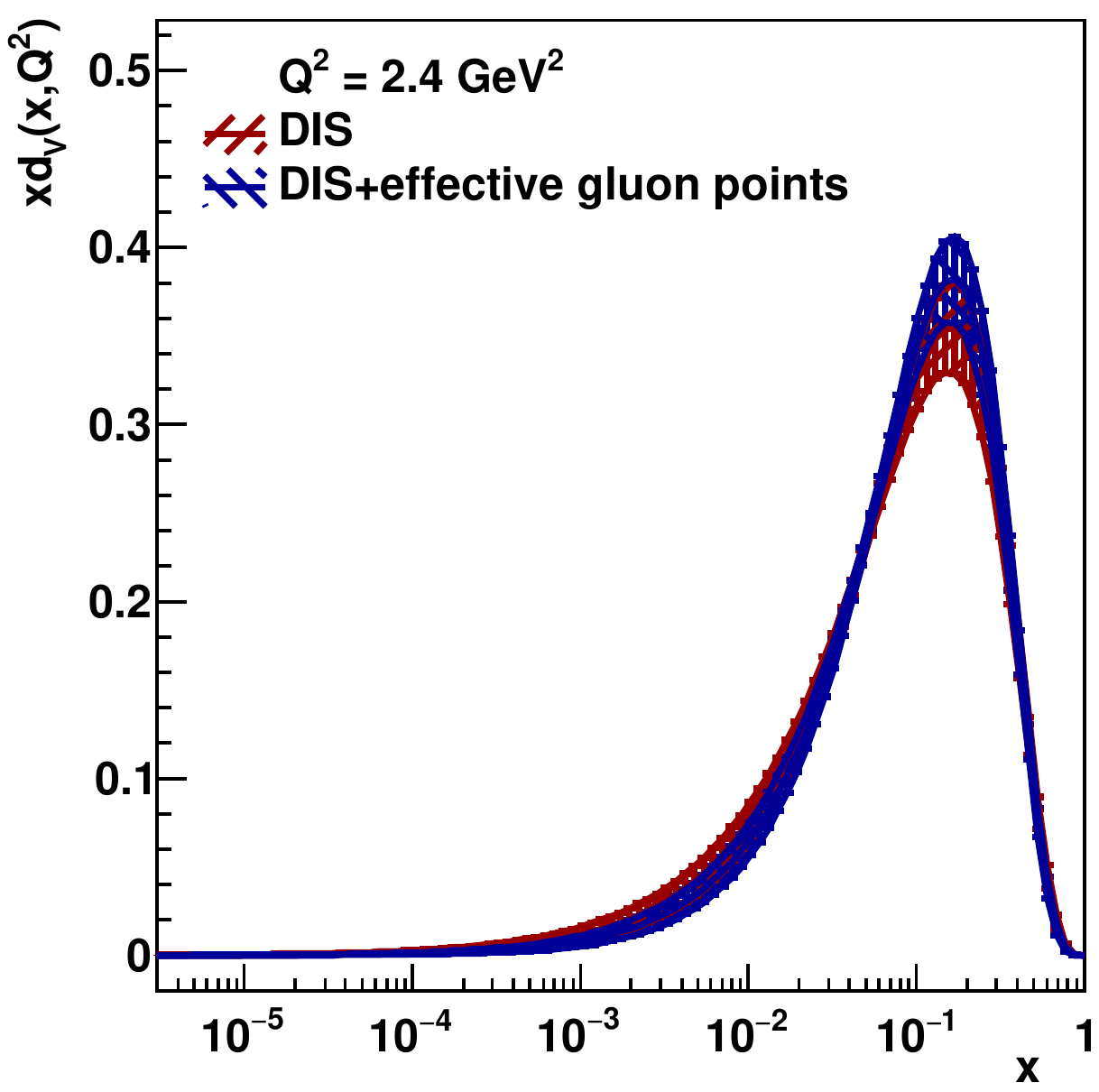}
\includegraphics[scale=0.38]{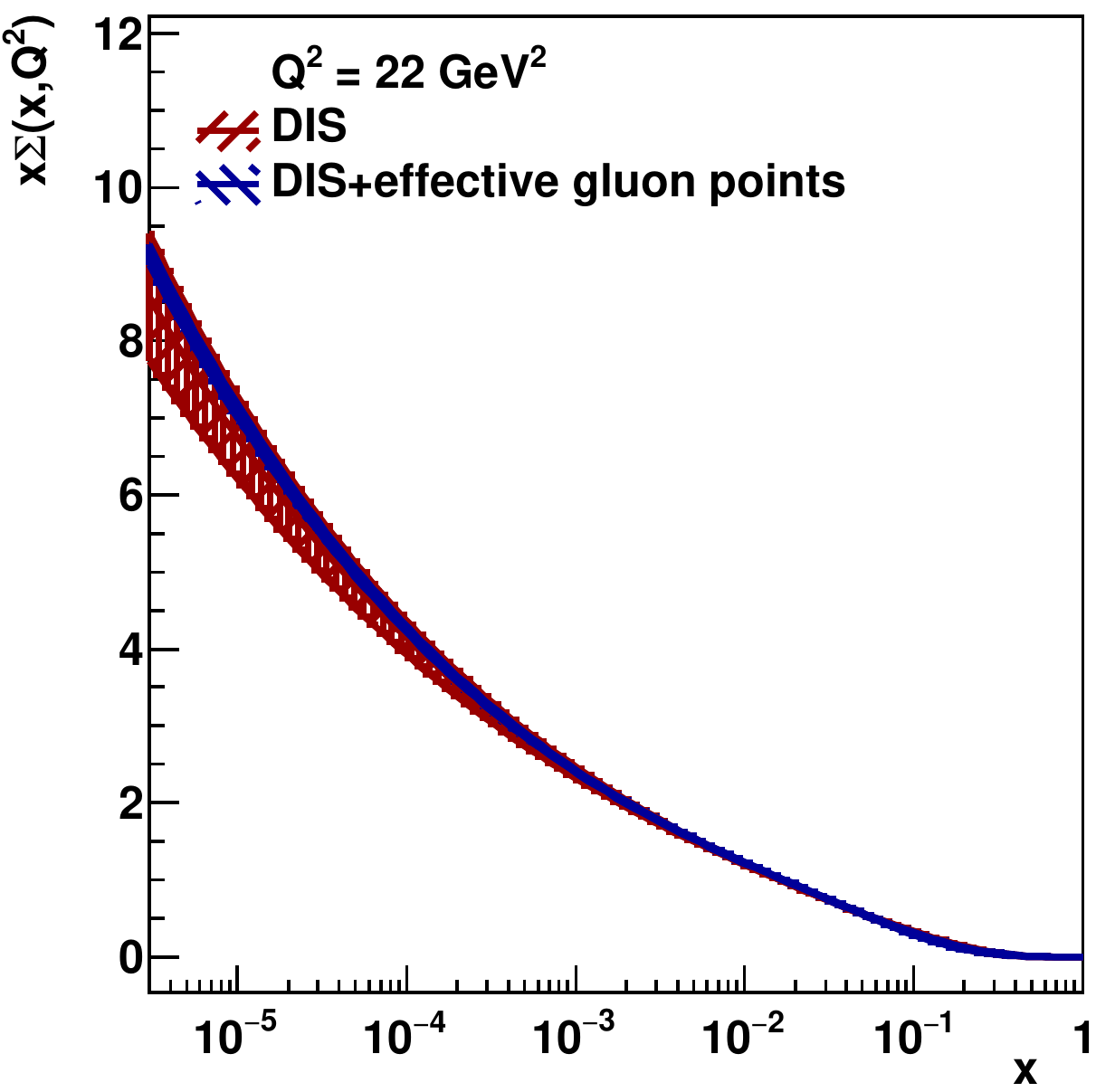}
\caption{\sf{The valence $xd_v$ quark distribution at scale 2.4 GeV$^2$ (left) and the singlet $x\Sigma$ quark distribution at 22  GeV$^2$ (right) obtained in \texttt{xFitter} fitting DIS data (red) and DIS+effective gluon data points~(blue).}}
\label{fig:5}
\end{center}
\end{figure}

It is clearly seen in Fig.~4 that the inclusion of the low-$x$ effective gluon points extracted from the exclusive heavy vector-meson production at the LHC essentially improve the accuracy of the global parton analysis, and we obtain a good overall final $\chi_{\text{min}}^2/\text{d.o.f} \sim O(1)$. The blue error band is much smaller than the red one.%
\footnote{As usual, additional theoretical uncertainties such as the correction to the Non-Relativistic QCD (NRQCD) approach used for the
$q\bar q \to V$ transition in the computation of $\sigma(\gamma p\to  V+p)$, or possible higher twist effects, or effects of absorptive corrections,
or other higher-order corrections, are not shown.
Note that, apart from the absorptive effects, all other corrections (e.g. NRQCD) do not affect the energy behaviour of the amplitude, but just change its normalisation. Indeed, to `know' the initial energy we need an additional gluon, which connects the upper (photon block in Fig. 1) with the lower/target parts of the diagram. This would correspond to a higher-twist contribution and should be suppressed by a large factorisation scale $\mu_F$. Typically, these corrections act `locally' in rapidity space and therefore do not affect the energy behaviour of the amplitude, only its normalisation.
On the other hand, our approach, in its present form, satisfactorily describes the available HERA data~\cite{prd102}. Recall that, as was shown in~\cite{Hoodbhoy:1996zg}, normalising the vector-meson wave function to the experimentally measured $J/\psi \to l^+l^-$ decay width allows us to account for relativistic corrections in the NRQCD approach with good accuracy (within a few percent).}
Moreover, the inclusion of new information obtained from the heavy vector-meson ultraperipheral production affects (albeit expectedly not so strongly) even the quark distribution as demonstrated, for example, in Fig.~\ref{fig:5}. The new $g_{\text{eff}}$ low-$x$ points affect the quark distribution even at a rather large $x$ due to the energy-momentum sum rules and the particular choice of the fixed form of the input ansatz.

\begin{table}[ht]
\centering
\begin{tabular}[t]{lcccc}
\toprule
Dataset&$\chi_{\text{min}}^2/\text{d.o.f}$~(\text{DIS})&$\chi_{\text{min}}^2/\text{d.o.f}$~(\text{DIS+eff. gluon pts.})&\\
\midrule
HERA1+2 NCep 820&80/73&79/73\\
HERA1+2 NCep 460&220/207&220/207\\
HERA1+2 CCep&43/39&44/39\\
HERA1+2 NCem&221/159&220/159\\
HERA1+2 CCem&54/42&56/42\\
HERA1+2 NCep 575&223/257&227/257\\
HERA1+2 NCep 920&465/391&470/391\\
LHC excl. $J/\psi$ $pp$ 7 TeV&N/A& 8.95/10\\
LHC excl. $J/\psi$ $pp$ 13 TeV&N/A& 3.51/10\\
LHC excl. $\Upsilon$ $pp$ 7,8 TeV&N/A& 3.23/3\\
\midrule
Total $\chi_{\text{min}}^2/\text{d.o.f}$&1412/1154 $\approx$ 1.22& 1444/1177 $\approx$ 1.23\\
\bottomrule
\end{tabular}
\caption{\sf{The partial $\chi_{\text{min}}^2/\text{d.o.f}$ for each dataset included in the baseline fit (DIS) and with the effective gluon data points added (DIS + eff. gluon pts.). Here, we use the DIS data in the kinematic range $Q^2 > 2.4$ GeV$^2$. The total $\chi_{\text{min}}^2/\text{d.o.f}$ is also given.}} 
\label{tab:3}
\end{table}

\section{Working to NNLO via a $K$ factor}
In principle, the method of ``effective gluon points'' described above can be extended to NNLO. Unfortunately at present the coefficient functions for $J/\psi$ and $\Upsilon$ photo- and electroproduction are known only at NLO level~\cite{nlo,sthesis,Flett:2021ghh}.

However, it is possible to include these points in a NNLO global analysis in an approximate way by introducing a NNLO/NLO $K$ factor, whose value can be extracted from the description of existing HERA $\gamma+p\to J/\psi+p$  (and/or $\gamma+p\to \Upsilon+p$) data. In the following, we denote this analysis NNLO* to distinguish it from an analysis in which the full NNLO coefficient functions would be used when they become available. In general the $K$ factor may depend on the factorization scale and on the ratio $z=(X+\xi)/2\xi$ of the parton momentum fractions $x=(X+\xi)$ to $\xi$.  
That is the $K$ factor should be included in the convolution for the $\gamma+p\to J/\psi+p$ amplitude
\be
\label{fact}
{\cal M}(\gamma+p\to J/\psi+p)~~=~~\sum_{i=g,q} N_i\int_{-1}^1 \frac{dX}X ~F_i(X,\xi)~K ~C^{{\rm NLO}}_{i}(\xi/X)\ ,
\ee
where the constants $N_i$ provide the correct normalization, see e.g.~\cite{prd102}, and $F_i$ and $C_i$ denote quark and gluon GPDs and coefficient functions respectively.
In our case the factorization scale is fixed, $\mu_F=M_{J/\psi}/2$, and the four-momentum transfer $t=t_{\rm min}=(2\xi m_p)^2/(1-2\xi)\sim 0$,
so 
these parameters are not shown explicitly in eqn.~(8).
The integral in eqn.~\eqref{fact} converges\footnote{It is known that the $\gamma+p\to J/\psi+p$ amplitude increases as a power of the photon energy (i.e.\ as a power of $1/\xi$). On the other hand the NNLO gluon coefficient function can contain only $\ln^2z$ from the higher-order gluon loop insertions. Thus at large energy (i.e.\ very small $\xi$)   the dominant contribution to (\ref{fact}) comes from $z\sim O(1)$ (i.e.\ $X\sim \xi$) while the power growth of the amplitude is provided by the GPD power behaviour. In this case we have to calculate the $\gamma+p\to J/\psi+p$ NNLO amplitude, ${\cal M}$, by multiplying the NLO coefficient functions, $C_i{^{\rm NLO}}$ with $i=g,q$, by this $K$ factor.}
at $X\sim \xi$. This means that actually after the convolution of the coefficient function with the gluon distribution the effective NNLO/NLO $K$ factor 
may be considered as a 
constant
{\em assuming that the power $\lambda$ of eqn.~(\ref{gg})  does not vary in the interval of interest}.

Thus the value of this $K$ factor can be estimated by describing the exclusive $J/\psi$ photoproduction data from HERA with the NNLO partons and NLO coefficient functions in the region $0.01>x>0.001$ since here the uncertainties in the gluon distribution given by the present global analysis are relatively small. 
In other words the square of the $K$ factor can be obtained as the ratio of the measured $\gamma+p\to J/\psi+p$ cross section, $\sigma({\rm data})$, to $\sigma^{{\rm NLO}/{\rm NNLO}}$ calculated using the NLO coefficient function but NNLO input partons (PDF/GPD). That is, 
\be
K_i=\sqrt{\left(\frac{\sigma({\rm data})}{\sigma^{{\rm NLO}/{\rm NNLO}}}\right)_i}\,,
\ee 
and then make the trivial average of the $K$ factors for each data point. From Fig.~\ref{fig:3} we may hope that the power $\lambda$ which describes the energy behaviour of the $\gamma+p\to J/\psi+p$ cross section is more or less constant. However, as it is seen from eqn.~\eqref{fact}, this would require the pure power behaviour for the gluons and not for the $J/\psi$ cross section. Unfortunately in the present analysis just in the region of interest $0.01>x>0.001$ we observe that the gluon PDF is too flat (see Fig.~\ref{fig:4} left). The power growth starts only at $x<0.001$.
Most probably this is explained by the fact that the DGLAP evolution does not account for the (higher twist) absorptive corrections which are not negligible in this  low-$x$ region. To mimic the role of these corrections in DIS data the fit chooses gluons which  slightly decrease at $x<0.01$.

In a future analysis, we have two possibilities. Either to include these absorptive effects into the DGLAP evolution following the GLR-MQ scheme~\cite{GLR,MQ} (with the possibility that this will generate power increasing gluons already in the HERA domain, $x<0.01$), or to have at hand the full NNLO coefficient function to reach complete NNLO accuracy and go beyond the NNLO* approach discussed here.

In any case in order to obtain realistic partons at such low $x$ and scales, the absorptive effect should be accounted for.

\section{Conclusion and outlook} 
We have described the method of ``effective gluon points'' which readily allows for the inclusion of exclusive heavy vector-meson $V$ photoproduction data, where $V = J/\psi, \Upsilon$ (as well as the exclusive production in ultraperipheral events at the LHC) in a conventional global parton analysis. Using \texttt{xFitter}, we fit DIS data together with these effective gluon points extracted from the $J/\psi$ and $\Upsilon$ exclusive production data from LHCb and
demonstrate that this crucially improves the accuracy of the obtained gluon distribution in the low-$x$ domain. The values of $xg_{\rm eff}(x)$ extracted from the available exclusive $J/\psi$ and $\Upsilon$ data from LHCb are presented in Tables~\ref{tab:1} and \ref{tab:2}. 

Since at present the NNLO coefficient functions
for the photon to vector meson ($\gamma\to V$) transition are not available, our computations were performed at NLO level. However, there is the possibility to reach
NNLO* accuracy,
by extracting the NNLO/NLO $K$ factor from the existing HERA $\gamma+p\to J/\psi+p$ data as discussed in Section~4. Moreover, exclusive vector-meson production data have already been collected by the LHC and more should follow.  
This merits further theoretical study of the NNLO* analysis, which we reserve for future work. 

The analysis presented in this work has demonstrated the potential of existing exclusive data to significantly improve the NLO global parton distributions at small $x$.

\section{Acknowledgements}
This project has received funding from the European Union’s Horizon 2020 research and innovation programme under grant agreement No. 824093 in order to contribute to the EU Virtual Access ``NLOAccess". This project has also received funding from the Agence Nationale de la Recherche (ANR) via the grant ANR-20-CE31-0015 (``PrecisOnium”) and via the IDEX Paris-Saclay ``Investissements d’Avenir” (ANR-11-IDEX-0003-01) through the GLUODYNAMICS project funded by the ``P2IO LabEx (ANR-10-LABX-0038)”. This work was also partly supported by the French CNRS via the IN2P3 Master Project ``QCDFactorisation@NLO".
The work of TT was supported by the STFC Consolidated Grant ST/T000988/1 and currently by ST/X000699/1.

\end{document}